\documentstyle[12pt,epsfig]{article}

\newlength{\dinwidth}
\newlength{\dinmargin}
\def\lapproxeq{\lower .7ex\hbox{$\;\stackrel{\textstyle
<}{\sim}\;$}}
\def\gapproxeq{\lower .7ex\hbox{$\;\stackrel{\textstyle
>}{\sim}\;$}}

\def\be{\begin{equation}}
\def\ee{\end{equation}}
\def\bea{\begin{eqnarray}}
\def\eea{\end{eqnarray}}

\def\funp{{I\!\!P}}

\begin{document}
\titlepage
\begin{flushright}
IPPP/02/03 \\
DCPT/02/04 \\
4 April 2002 \\
\end{flushright}

\vspace*{2cm}

\begin{center}
{\Large \bf Photon-exchange processes at hadron colliders as a}

\vspace*{0.5cm}

{\Large \bf probe of the dynamics of diffraction}

\vspace*{1cm}
V.A. Khoze$^{a,b}$, A.D. Martin$^a$,  and M.G. Ryskin$^{a,b}$ \\

\vspace*{0.5cm} $^a$ Department of Physics and Institute for
Particle Physics Phenomenology, University of
Durham, Durham, DH1 3LE \\
$^b$ Petersburg Nuclear Physics Institute, Gatchina,
St.~Petersburg, 188300, Russia
\end{center}

\vspace*{1cm}

\begin{abstract}
The rich structure of photon-exchange processes at hadron
colliders is studied.  We discuss central vector meson production
$(pp \rightarrow p + J/\psi + p)$, $W$ production $(pp \rightarrow
p + W + X)$ and $\mu^+ \mu^-$ production.  Each process has
distinct, and large, soft $pp$ rescattering effects, which can be
directly observed by detecting the outgoing protons.  This allows
a probe of the optical density of the proton, which plays a
crucial role in the evaluation of the rapidity gap survival
probabilities in diffractive-like processes at hadron colliders.  We
note that an alternative mechanism for $J/\psi$ production is
odderon, instead of photon, exchange; and that the ratio of
odderon to photon contributions is enhanced (suppressed) for $\phi
(\Upsilon)$ vector meson production.
\end{abstract}

\newpage
\section{Introduction}

Processes with rapidity gaps at hadron colliders provide an
attractive possibility (i) to search for New Physics (Higgs boson,
SUSY particles, etc.) in a clean environment (see for example, \cite{KMR}
and references therein) and (ii)
to study the properties of the diffractive amplitude.
Unfortunately, the cross sections for such processes are suppressed
by the small probability, $\hat{S}^2$, that the rapidity gaps
survive soft rescattering effects between the interacting hadrons,
which can generate secondary particles populating the gaps
\cite{DKS}--\cite{KKMR}.

In general, we may write the survival factor $\hat{S}^2$ in a
multi-channel eikonal framework in the form
\begin{equation}
\label{eq:a1}
 \hat{S}^2 \; = \; \frac{\int \sum_i \left |{\cal M}_i (s, b_t^2)
 \right |^2 \: \exp \left (- \Omega_i (s, b_t^2) \right ) d^2 b_t}{\int
 \sum_i \left | {\cal M}_i (s, b_t^2) \right |^2 d^2 b_t}
\end{equation}
where an incoming proton is decomposed into diffractive
eigenstates, each with its own opacity\footnote{Really we deal
with a matrix $\Omega_{jj^\prime}^{ii^\prime}$, where the indices
refer to the eigenstates of the two incoming and two outgoing
hadrons. \cite{KKMR}} $\Omega_i$. Here ${\cal M}_i (s,b_t^2)$ are the amplitudes
(in impact parameter $b_t$ space) of the process of interest at
centre-of-mass energy $\sqrt{s}$. They may be different
for the different diffractive eigenstates. It is important
to note that the suppression factor $\hat{S}^2$ is not universal,
but depends on the particular hard subprocess, as well as on the
kinematical configurations of the parent reaction \cite{KKMR}.

The opacities $\Omega_i (s, b_t)$ of the proton have been
calculated in a number of models
\cite{GLM,KMRhiggs,KMRsoft,BH,KKMR}, and used to determine
$\hat{S}^2$ for various rapidity gap processes.  However it is
difficult to guarantee the precision of predictions which rely on
soft physics.  The calculations of $\hat{S}^2$ can, in principle,
be checked by computing the event rate for processes such as
central $Z$ production by $WW$ fusion \cite{CZ} or central dijet production
with a rapidity gap on either side \cite{KMR00,KMRhiggs}, and comparing with the
experimental rate.  However, to date, the only such check has been
the prediction of the diffractive dijet production at the Tevatron
in terms of the diffractive structure functions measured at HERA
\cite{KKMR}.  We may regard these as `integrated' checks.\footnote{Another
probe of the models for soft diffraction comes from the comparison of the
experimental upper limit on the exclusive dijet production rate with the
theoretical expectation \cite{KMR01}.} It is
clearly desirable to study the {\it profile} of the optical
density $\Omega (s, b_t^2)$ itself for some well known {\it
observable} reaction, rather than simply the integrated quantity
$\hat{S}^2$.

Here we show that central production processes mediated by photon exchange offer an
excellent possibility to probe the opacity $\Omega (s, b_t^2)$ in
more detail.  The generic diagram for such a process is shown in
Fig.~1(a), together with the definition of the kinematic
variables.  For very small photon momentum transfer $t = q_1^2$,
the scattering occurs at large impact parameters, outside the strong
interaction radius, where the opacity is essentially zero, and
$\hat{S}^2 \simeq 1$.  As $|t|$ increases we probe smaller and
smaller $b_t$ and the opacity increases.  In this way we can {\it
scan} the opacity $\Omega (s, b_t^2)$.  In terms of Feynman
diagrams the rescattering induces interference
between the contributions from Figs.~1(a) and 1(b).  Effectively,
the $1/t$ form of the photon exchange amplitude is replaced by a
more complicated form with a diffractive minimum in the region
$|t| \sim 0.1~{\rm GeV}^2$.  To obtain the cleanest probe of this
effect it is best to consider a reaction where the whole impact
parameter distribution is driven by the photon propagator, while
the amplitude for the $\gamma p$ subprocess samples a concentrated
region, $\Delta b_t$, of the impact parameter space.

\begin{figure}[!h]
\begin{center}
\epsfig{figure=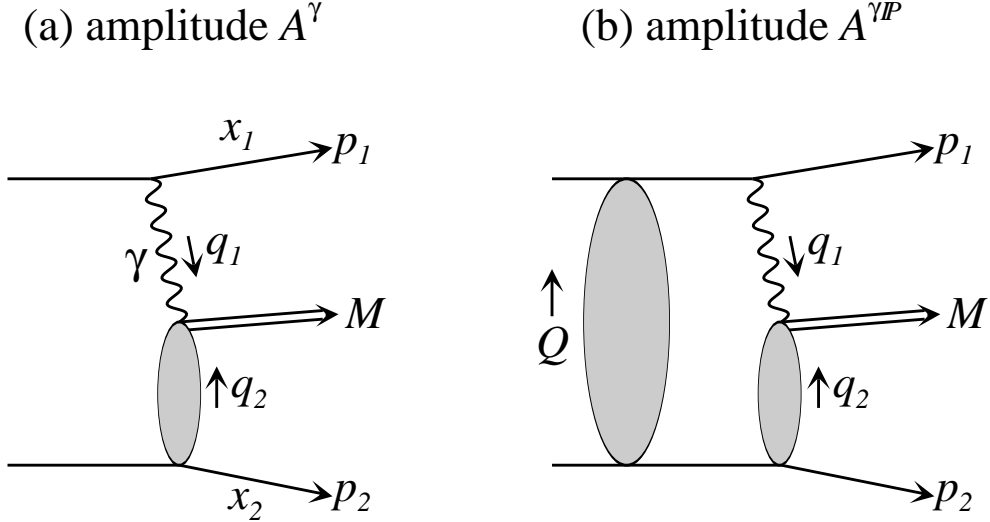,height=3in}
 \caption{The exclusive $pp \rightarrow p + M + p$ process mediated by
 photon exchange (a) without and (b) with rescattering corrections.  The
 particle four momenta are indicated.  In this paper we take the system
 $M$ to be the $J/\psi$ vector meson, the $W$ boson and, finally, a $\mu^+ \mu^-$ pair.
 For $W$ production we have, of necessity, proton dissociation at the lower vertex; that is
 $pp \rightarrow p + W + X$.}
 \label{Fig1}
\end{center}
\end{figure}

Moreover, we must study a process where photon exchange is a major
contribution, and the background mechanisms are relatively small.
One possibility is to observe the exclusive reaction
\begin{equation}
\label{eq:a2}
 pp \; \rightarrow \; p \: + \: J/\psi \: + \: p,
\end{equation}
where the $+$ sign indicates the presence of a rapidity gap.  To
reduce the spread of $\Delta b_t$ we should select events with
relatively large momentum transfer $(p_{2t} \sim {\cal O}(1$~GeV)) in
the quasi-elastic subprocess
$\gamma p \rightarrow J/\psi + p$.  Other examples are central $W$
boson production via $\gamma W$ fusion and central $\mu^+ \mu^-$
production via $\gamma\gamma$ fusion.  That is the system $M$ in
Fig.~1 is either the $J/\psi$ or the $W$ boson or a massive $\mu^+
\mu^-$ pair.

\section{Rescattering in $\gamma$ exchange proceses:  a first look}

To gain insight into how $\gamma$ exchange processes allow a probe
of the rapidity gap survival factor, we first perform a simplified
evaluation of the diagrams of Fig.~1.  As mentioned above we are
interested in the regime where $p_{1t} \simeq 0-0.4$~GeV and
$p_{1t}^2 \ll p_{2t}^2$.  (One consequence is that the diagram
with photon emission from the lower vertex is supressed.)~~To fix
the $\gamma p \rightarrow M p$ subprocess energy $W$ we have to
measure the longitudinal momentum fraction $\xi_1 = 1 - x_1$
carried by the photon, where $x_1$ is that of the detected proton.

In the small $p_{1t}$ regime, the cross section, neglecting the
rescattering contribution Fig.~1(b), may be written in the
factorized form
\begin{equation}
\label{eq:a3}
 \sigma \; = \; \int \: dN (\xi_1) \: \sigma_{\gamma p \rightarrow
 Mp}(W),
\end{equation}
where $N (\xi_1)$, the effective number of photons, is well known \cite{bgms}.
For small $t$ it is safe to neglect longitudinally polarised
photons and to consider only the number of transversely polarised
photons. For $\xi_1 \ll 1$, neglecting terms of higher order in $\xi_1$ we have
\begin{equation}
\label{eq:a4}
 dN^T (\xi_1) \; = \; \frac{d^2q_{1t} q_{1t}^2}{(q_{1t}^2 + \xi_1^2
 m_N^2)^2} \: \frac{\alpha}{\pi^2} \: F_N^2 (t) \left (1 - \xi_1 +
 \frac{1}{2} \xi_1^2 \right ) \: \frac{d\xi_1}{\xi_1},
\end{equation}
where, in the absence of rescattering $\vec{q}_{1t} =
-\vec{p}_{1t}$.  Here $\alpha$ is the QED coupling and the
expression in brackets in the numerator is the QED splitting
function for emission of a transversely polarised photon from the
proton.  The photon propagator can be written as
\begin{equation}
\label{eq:a5}
 t \; = \; - \frac{\left (p_{1t}^2 \: + \: \xi_1^2 m_N^2 \right )}{1 \: - \: \xi_1},
\end{equation}
where $|t_{\rm min}| = \frac{\xi_1^2 m_N^2}{1-\xi_1}$ and $m_N$ is
the proton mass.  $F_N (t)$ is the electromagnetic form factor of
the proton, which, in the simplified discussion presented in this
section, we take to be $F_1 (t)$.  We may thus write the amplitude
for Fig.~1(a) in the form
\begin{equation}
\label{eq:a6}
 A^\gamma \; = \; {\cal M} \exp (-bq_{2t}^2/2) \: \left ( \frac{q_{1t}}{q_{1t}^2
 + \xi_1^2 m_N^2} \right ) \: F_1 (t),
\end{equation}
where $b$ is the slope of the differential cross section of the
$\gamma p \rightarrow Mp$ subprocess, and ${\cal M}$ contains the
$W$ dependence of the subprocess, as well as the remaining $\xi_1$
dependence and other factors in (\ref{eq:a4}).

To calculate the rescattering contribution, Fig.~1(b), we use the
momentum representation.  Throughout the paper we neglect the rescattering
of the system $M$, as it has a much smaller cross section.  We may
also neglect the spin flip component in the proton-Pomeron
vertex\footnote{This component is expected to be small and
consistent with zero.  If we note the similarity between the
photon and Pomeron vertices then the magnitude of the isosinglet
spin-flip amplitude is proportional to $\left |\frac{1}{2}
(\mu_p^a + \mu_n^a) \right | \lapproxeq 0.06$, where the anomalous
magnetic moments $\mu^a$ of the neutron and proton cancel each
other almost exactly.}.  To estimate the qualitative features of
the rescattering effect we assume, in this Section, that the
amplitude for elastic proton-proton scattering, at energy
$\sqrt{s}$ and momentum transfer $k_t$, has the simplified form
\begin{equation}
\label{eq:a7}
 A_{pp} (s, k_t^2) \; = \; A_0 (s) \: \exp (-B k_t^2/2).
\end{equation}
>From the optical theorem we have ${\rm Im} A_0 (s) = s
\sigma_{pp}^{\rm tot} (s)$, and for the small contribution of the
real part it is sufficient to use ${\rm Re} A_0/{\rm Im} A_0
\simeq 0.13$ in the energy regime of interest.  $B$ is the slope
of the elastic $pp$ differential cross section, $d \sigma_{pp}/dt
\propto \exp (Bt)$.

Using the above elastic $pp$ amplitude we may write the
rescattering contribution, Fig.~1(b), to the $pp \rightarrow p + M
+ p$ amplitude as
\begin{equation}
\label{eq:a8}
 A^{\gamma \funp} \; = \; i \int \frac{d^2 Q_t}{8 \pi^2} \left (
 \frac{q_{1t} F_1 (t)}{q_{1t}^2 + \xi_1^2 m_N^2} \right ) \: {\cal
 M} \exp (-bq_{2t}^2/2 - BQ_t^2/2) \: \frac{A_0 (s)}{s}
\end{equation}
with
\begin{equation}
\label{eq:a9}
 \vec{q}_{1t} \; = \; \vec{Q}_{t} - \vec{p}_{1t}, \quad\quad
 -\vec{q}_{2t} \; = \; \vec{Q}_{t} + \vec{p}_{2t}.
\end{equation}
The photon induced singularity ($\sim 1/q_{1t}$ for $\xi_1
\rightarrow 0$) is integrable in (\ref{eq:a8}).  The main
contribution comes from the region $Q_t^2 \sim 2/B$.  For small
$p_{1t}^2 \ll (p_{2t}^2, 2/B)$, the $Q_t$ integration may be
performed, and we find
\begin{equation}
\label{eq:a10}
 A^{\gamma \funp} \; \sim \; i \: \frac{A_0 (s)}{8 \pi s} {\cal M}
 \sqrt{\frac{2 \pi}{(B + b)}} ~~\exp \left ( - \: \frac{bB}{2 (b +
 B)}\: p_{2t}^2 \right ).
\end{equation}
\begin{figure}[!h]
\begin{center}
\epsfig{figure=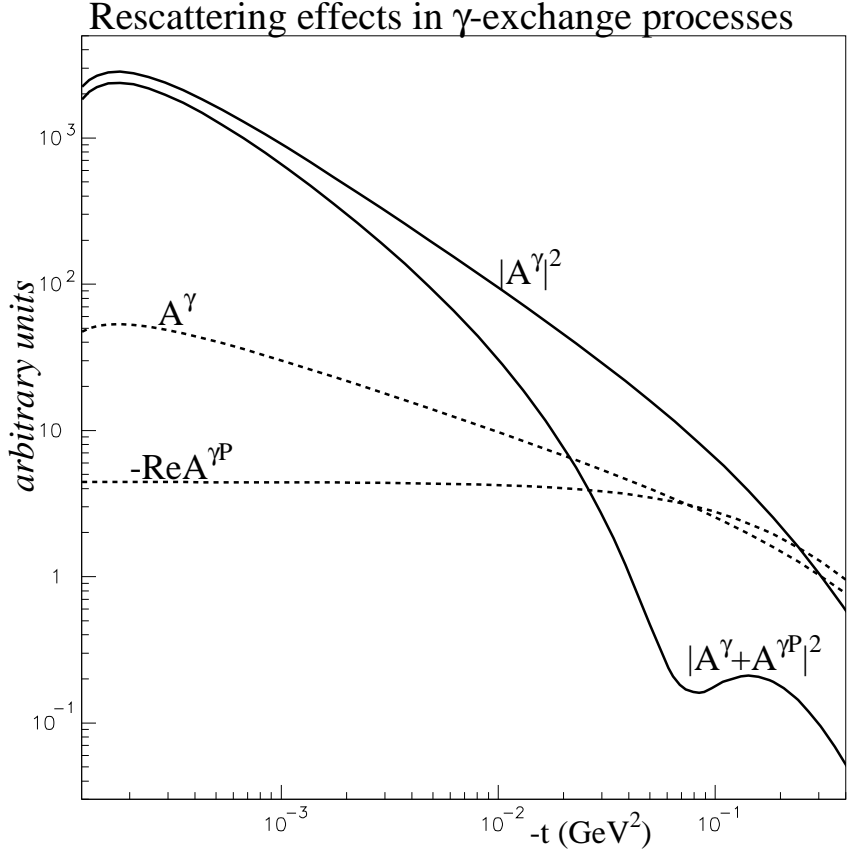,height=5in}
 \caption{The cross sections, $|A^\gamma|^2$ and $|A^\gamma + A^{\gamma \funp}|^2$,
 for the $\gamma$-exchange process $pp \rightarrow p + M + p$ of Fig.~1, without and
 with rescattering corrections respectively.  The dashed curves show the contributing
 amplitudes calculated from (\ref{eq:a6}) and (\ref{eq:a8}), in the simple case with
 slope $b = 0$.  For the elastic proton-proton amplitude, $A_{pp}$, we take (\ref{eq:a7})
 with slope $B = 17~{\rm GeV}^{-2}$ and $\sigma_{pp}^{\rm tot} = 76$~mb, which correspond
 to the Tevatron measurements \cite{CDF}.}
 \label{Fig2}
\end{center}
\end{figure}

The effect of rescattering is shown in Fig.~2.  Here, for
simplicity, we have set the slope $b = 0$, that is the $\gamma p
\rightarrow Mp$ subprocess is assumed to occur at zero impact
parameter.  The upper curve, $|A^\gamma|^2$, is proportional to
the cross section for the photon-mediated process $pp \rightarrow
p + M + p$ of Fig.~1(a) in the absence of rescattering, whereas
the $|A^\gamma + A^{\gamma \funp}|^2$ curve shows the effect of
including the rescattering corrections of Fig.~1(b).  The
diffractive dip in the region of $-t \sim 0.1~{\rm GeV}^2$, due to
the destructive interference of the $A^\gamma$ and $A^{\gamma
\funp}$ amplitudes, is clearly evident.  The dip is partially
filled in by the presence of the ${\rm Re} A_0$ contribution
to (\ref{eq:a7}), which leads to a small ${\rm Im} A^{\gamma
\funp}$ term.  It is clear that a measurement of the $t$
dependence of the cross section will provide a scan of the
rapidity gap survival probability $\hat{S}^2$, that is of the proton
optical density $\Omega(s,b_t^2)$. It is seen that
rescattering gives more than an order of magnitude suppression in
the region $0.03 \lapproxeq -t \lapproxeq 0.2~{\rm GeV}^2$.

So far we have considered only the amplitude which conserves the
$s$-channel proton helicity at the vertex with the exchanged
photon.  This non-flip amplitude is described by the $F_1 (t)$
electromagnetic form factor of the proton.  However as $-t$
increases the helicity at the photon vertex may be flipped by the
proton's anomalous magnetic moment.  The spin-flip amplitude is
described by the form factor $F_2 (t)$.  Neglecting terms of order
$\xi_1^2$ the amplitude is given by an expression analogous to
(\ref{eq:a6}), with $F_1 (t)$ replaced by $(q_{1t}/2m_N) F_2 (t)$.
In almost all of the results presented below we include the
spin-flip contribution, although, for simplicity, we write the
formulae in terms of the non-flip amplitude alone.  The exception
is Fig.~3. The comparison of Fig.~3 with Fig.~4 shows that the
effect of the spin-flip contribution is quite small.

\section{Second look:  inclusion of photon polarisation}

So far we have studied photon-exchange reactions in hadron
colliders with the neglect of photon polarisation, as is often
done in the equivalent photon approach.
However, when we include soft $pp$ rescattering effects, we must
take care. Frequently for photon exchange processes we deal with
expressions of the form
\begin{equation}
\label{eq:a11}
 |A|^2 \; = \; \vec{\varepsilon} \cdot \vec{\varepsilon}^{~\prime
 *} |T|^2,
\end{equation}
where $\vec{\varepsilon}$ and $\vec{\varepsilon}^{~\prime *}$ are
the photon polarisation vectors of the amplitude and the complex
conjugate amplitude respectively.  At small $\xi_1$ the
polarisation vector $\vec{\varepsilon}$ is aligned with the
transverse momentum $\vec{q}_{1t}$.  We take $\vec{\varepsilon}$
in the direction $- \vec{q}_{1t}$.  Without rescattering,
Fig.~1(a), we have $\vec{q}_{1t} = -\vec{p}_{1t}$, whereas for the
rescattering contribution, Fig.~1(b), we have $\vec{q}_{1t} =
\vec{Q}_t - \vec{p}_{1t}$.  In the latter case we see that the
photon polarisation depends on the loop momentum $\vec{Q}_t$, and
so we must work in terms of the two different polarisation states.
It is convenient to choose a linear polarisation basis. We take
one polarisation vector $\vec{\varepsilon}_1$ aligned with
$\vec{p}_{1t}$ (which is the polarisation vector of the photon in
the absence of rescattering), and the other,
$\vec{\varepsilon}_2$, perpendicular to $\vec{\varepsilon}_1$.
The amplitude $A^\gamma$ of Fig.~1(a) proceeds only via the first
polarisation vector and hence the $pp \rightarrow p + M + p$ cross
section via $\gamma$-exchange has the form
\begin{equation}
\label{eq:a12}
 \sigma \; \propto \; |A^\gamma + A_1^{\gamma \funp} |^2 \: + \:
 |A_2^{\gamma \funp}|^2.
\end{equation}

Let us study the impact of $\vec{\varepsilon} \cdot
\vec{\varepsilon}^{~\prime *}$ on the interference term between
the amplitudes $A^\gamma$ and $A^{\gamma \funp}$.  In general, we
have
\begin{equation}
\label{eq:b12}
 A^\gamma A^{\gamma \funp *} \; \sim \; - \vec{\varepsilon}_1 \cdot
 \vec{\varepsilon}^{~\prime *} \; \sim \; - \vec{p}_{1t} \cdot
 (\vec{p}_{1t} - \vec{Q}_t),
\end{equation}
where the minus sign arises because $A^{\gamma \funp}$ is negative
relative to $A^\gamma$ due to the absorptive nature of Pomeron
exchange.  For very small values of $p_{1t}$ we see that
$\vec{\varepsilon}^{~\prime}$ is antiparallel to $\vec{Q}_t$.
Moreover, the main contribution of Fig.~1(b) comes from the
smaller values of $q_{2t}$, and so $\vec{Q}_t$ tends to be
antiparallel to $\vec{p}_{2t}$.  Hence
$\vec{\varepsilon}^{~\prime}$ tends to be parallel to
$\vec{p}_{2t}$, and
\begin{equation}
\label{eq:a13}
 A^\gamma A^{\gamma \funp *} \; \sim \; - \vec{\varepsilon}_1 \cdot
 \vec{\varepsilon}^{~\prime *} \; \sim \; - \vec{p}_{1t} \cdot
 \vec{p}_{2t} \quad\quad ({\rm for~very~small}~p_{1t}).
\end{equation}
On the other hand, as $p_{1t}$ increases,
$\vec{\varepsilon}^{~\prime}$ becomes more and more parallel to
$\vec{p}_{1t}$.  Hence
\begin{equation}
\label{eq:b13}
 A^\gamma A^{\gamma \funp *} \; \sim \; - \vec{\varepsilon}_1 \cdot
 \vec{\varepsilon}^{~ \prime *} \; \sim \; - \vec{p}_{1t} \cdot
 \vec{p}_{1t} \quad\quad ({\rm for~larger}~p_{1t}).
\end{equation}
>From (\ref{eq:a13}) we see that the interference term depends on
the azimuthal angle $\phi$ between the transverse momenta of the
outgoing protons, $\vec{p}_{1t}$ and $\vec{p}_{2t}$.  For very
small $\vec{p}_{1t}$, we have {\it constructive} interference if
$\vec{p}_{1t}$ and $\vec{p}_{2t}$ are back-to-back (that is if
$\phi = 180^\circ$), and {\it destructive} interference when
$\vec{p}_{1t}$ and $\vec{p}_{2t}$ are aligned ($\phi = 0$).  As
$p_{1t}$ increases, $\vec{\varepsilon}^{~\prime}$ becomes more and
more aligned with $\vec{p}_{1t}$, rather than $\vec{p}_{2t}$, and
we have destructive interference for all azimuthal angles $\phi$,
see (\ref{eq:b13}).

The evaluation of the exclusive process $pp \rightarrow
p + J/\psi + p$ is described in Section~5.  However it is
informative to show a sample of these results now, in order to
illustrate the effects of including the proper treatment of the
photon polarisation.  Fig.~3 shows the differential cross section
at energy $\sqrt{s} = 500$~GeV appropriate to RHIC for a realistic
choice of kinematic variables.  The (dotted) reference curve,
denoted by $|A^\gamma|^2$, is the naive estimate based on
Fig.~1(a).  It was obtained by multiplying the $\gamma p
\rightarrow J/\psi + p$ cross section by the equivalent photon
flux, (\ref{eq:a3}).  For simplicity here we neglect the
contributions which flip the spin of the proton.  The other curves
in Fig.~3 show the effect of including the rescattering
contribution of Fig.~1(b).  We take a realistic slope $b =
4.5~{\rm GeV}^{-2}$ for the $\gamma p \rightarrow J/\psi + p$
subprocess, as described in Section~5.

\begin{figure}[!h]
\begin{center}
\epsfig{figure=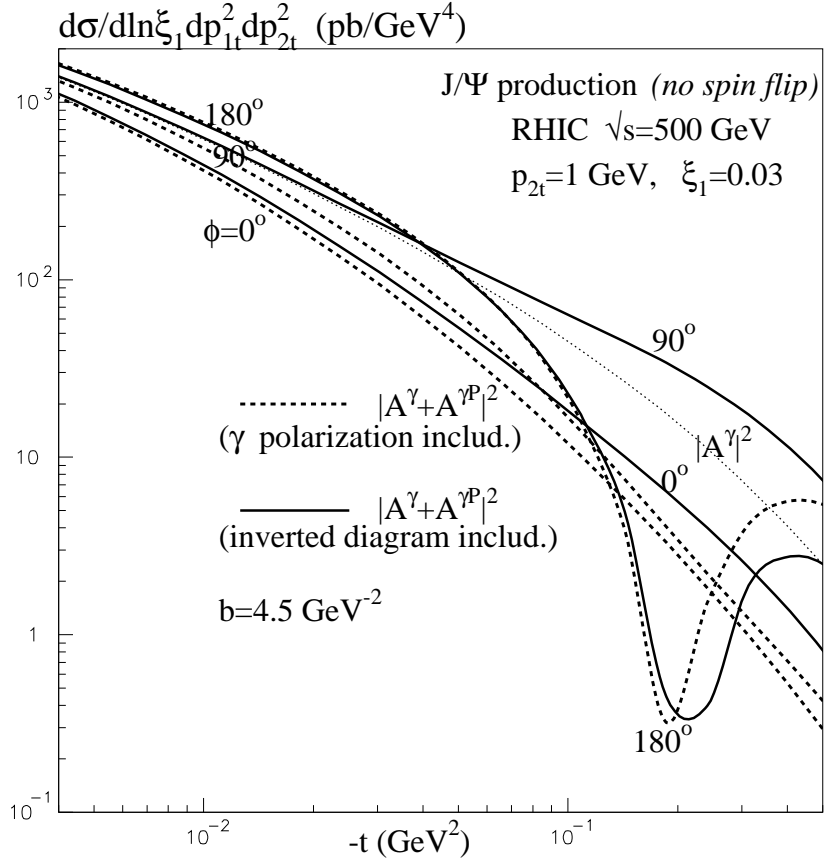,height=5in}
 \caption{The differential cross section for $pp \rightarrow p + J/\psi + p$
 via photon exchange at $\sqrt{s} = 500$~GeV.  The dashed curves include
 rescattering, with a proper treatment of photon polarisation.  The solid
 curves also include, besides the amplitudes of Fig.~1, the `inverted'
 amplitudes with the photon coupled to the lower proton.  The predictions
 are shown for three values of the azimuthal angle, $\phi = 0^\circ, 90^\circ,
 180^\circ$, between the transverse momenta $\vec{p}_{1t}$ and $\vec{p}_{2t}$
 of the outgoing protons. Here and in what follows $t$ is given by Eq.~(5).}
 \label{Fig3}
\end{center}
\end{figure}

The dashed curves in Fig.~3 show the predictions for the $pp
\rightarrow p + J/\psi + p$ cross section which include the
correct treatment of the photon polarisation, for three different
choices of the azimuthal angle, namely $\phi = 0, 90^\circ$ and
$180^\circ$. These results clearly demonstrate the anticipated
behaviour obtained in (\ref{eq:a13}) and (\ref{eq:b13}).  For
$\phi = 180^\circ$ the rescattering amplitude $A^{\gamma \funp}$
has the largest absolute value.  It is positive for small $-t$ and
reverses sign as $-t$ increases.  As a consequence, the diffractive
dip is deeper and narrower than the simple prediction shown in
Fig.~2.  For $\phi = 0$ and $90^\circ$ the amplitude $A^{\gamma
\funp}$ is negative everywhere and relatively smaller, which
shifts the respective dips outside the region of interest.

Besides, allowing for the effects of photon polarisation, there is
another complication that we must consider.  As $-t$ increases, we
have to include the contributions of diagrams analogous to those
shown in Fig.~1, but with the photon and Pomeron exchanges
interchanged.  That is we must include contributions from
`inverted' diagrams with the photon coupled to the lower proton
vertex.  Their contribution is negligibly small for very small
$q_{1t}$ and larger values of $\xi_1$, but increases significantly
in the dip region.  The results obtained after including these
extra amplitudes are shown by the three continuous curves in
Fig.~3.  It is interesting to note that for $\phi = 90^\circ$ the
extra diagrams have a large effect in the region where $p_{1t}$
becomes comparable to $p_{2t}$, and lead to a cross section even
larger than that due to the reference $|A^\gamma|^2$ prediction in
the absence of rescattering.

\section{Detailed formalism for rescattering corrections}

Before we consider specific examples of $pp \rightarrow p + M + p$
production via photon exchange, we describe a more realistic way
to evaluate rescattering corrections than that described in
Section~2.  We use the formalism and the results of
ref.~\cite{KMRsoft}.  The formalism embodies (i) pion-loop
insertions to the Pomeron trajectory, (ii) two-channel eikonal
description of proton-proton rescattering and (iii) high mass
diffractive dissociation.  The parameters of the model were tuned
to describe all the main features of the soft $pp$ data throughout
the CERN-ISR to the Tevatron energy interval.  In terms of the two-channel
eikonal the incoming proton is described by two
diffractive eigenstates $|\phi_i \rangle$, each with its own
absorptive cross section.

The eigenstates were taken to have the same profile in impact
parameter space, and absorptive cross sections
\begin{equation}
\label{eq:a14}
 \sigma_i \; = \; a_i \sigma_0 \quad\quad {\rm with} \quad\quad
 a_i \; = \; 1 \pm \gamma,
\end{equation}
where $\gamma = 0.4$ \cite{KMRsoft}.  That is the two channel
opacity is
\begin{equation}
\label{eq:a15}
 \Omega_{jj^\prime}^{ii^\prime} \; = \; \delta_{ii^\prime}
 \delta_{jj^\prime} a_i a_j \Omega.
\end{equation}
The impact parameter representation of the elastic amplitude is
thus
\begin{equation}
\label{eq:a16}
 \frac{1}{s} \: {\rm Im} \: \tilde{A}_{pp} (b_t) \; = \; \left ( 1 \:
 - \: \frac{1}{4} \left [e^{- (1 + \gamma)^2 \Omega/2} \: + \:
 2e^{- (1 - \gamma^2) \Omega/2} \: + \: e^{- (1 - \gamma)^2
 \Omega/2}\right ] \right ).
\end{equation}

As both the eigenstates $|\phi_i \rangle$ have the same $b_t$
profile, photon emission is controlled by the same proton
electromagnetic form factors $F_1 (t)$ and $F_2 (t)$, and there
are no off-diagonal transitions at the photon vertex
\begin{equation}
\label{eq:a17}
 \langle \phi_1 | \gamma | \phi_2 \rangle \; = \; \langle \phi_2 |
 \gamma | \phi_1 \rangle \; = \; 0.
\end{equation}
Therefore the amplitude of $pp$ rescattering, which occurs in
Fig.~1(b), takes the form
\begin{equation}
\label{eq:a18}
{\rm Im} \; \tilde{A}_{pp} (s, b_t) \; = \; s \left ( 1 \: - \: \frac{1}{4}
 \left [ (1 + \gamma) e^{- (1 + \gamma)^2 \Omega/2} \: + \: 2e^{-(1 -
 \gamma^2) \Omega/2}\: + \: (1 - \gamma) e^{- (1 - \gamma)^2
 \Omega/2}\right ] \right ).
\end{equation}
The extra $(1 \pm \gamma)$ factors reflect the different Pomeron
couplings to the $| \phi_i \rangle$ eigenstates in the $pp
\rightarrow p + M + p$ production amplitude, that is in the
right-hand part of Fig.~1(b).  The optical density $\Omega (s,
b_t)$ was given in Ref.~\cite{KMRsoft} for Tevatron $(\sqrt{s} =
2~{\rm TeV})$ and LHC $(\sqrt{s} = 14~{\rm TeV})$ energies.

As before, we work in momentum space, and replace (\ref{eq:a7}) by
\begin{equation}
\label{eq:a19}
 A_{pp} (s, k_t^2) \; = \; \frac{1}{2 \pi} \int d^2 b_t \:
 4 \pi \:  \tilde{A}_{pp} (s, b_t) \: e^{i \vec{k}_t \cdot \vec{b}_t}.
\end{equation}
Hence the amplitude of Fig.~1(b) becomes
\begin{equation}
\label{eq:a20}
 A^{\gamma \funp} \; = \; - \int \: \frac{d^2 Q_t}{8 \pi^2} \left
 ( \frac{q_{1t} F_1 (t)}{q_{1t}^2 + \xi_1^2 m_N^2} \right ) \: \frac{A_{pp} (s,
 Q_t^2)}{s} \: {\cal M} \: \exp (-b q_{2t}^2/2),
\end{equation}
in analogy to (\ref{eq:a8}).  The amplitude ${\cal M}$ is defined
below (\ref{eq:a6}).  Although here we show only Im~$A_{pp}$, we
include the contribution from Re~$A_{pp}$ as described in
Section~2.

\begin{figure}[!h]
\begin{center}
\epsfig{figure=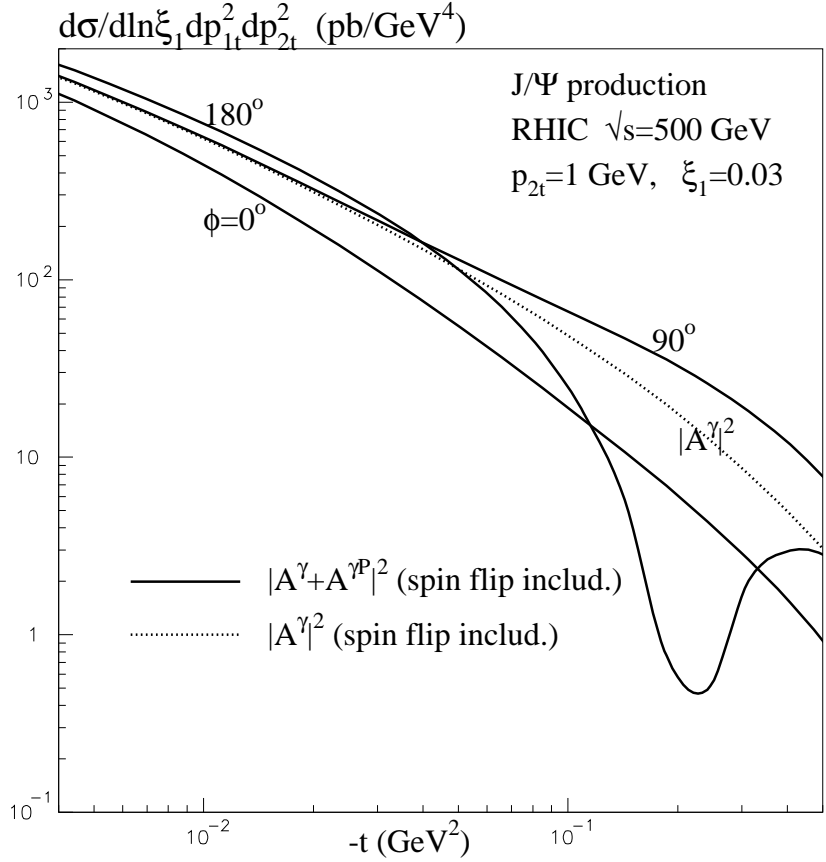,height=5in}
 \caption{The continuous curves are as in Fig.~3 but with the proton
 spin-flip amplitude included.}
 \label{Fig4}
\end{center}
\end{figure}

\section{Exclusive $J/\psi$ hadroproduction mediated by $\gamma$ exchange}

Here we study the photon-mediated exclusive reaction, $pp
\rightarrow p + J/\psi + p$ of Fig.~1, in more detail.  The cross
section of the $\gamma p \rightarrow J/\psi + p$ subprocess may be
calculated perturbatively (see, for example, \cite{R,RRML,FKS}) or
taken from extrapolations of the HERA data \cite{JPSI}.  The cross
section is well described by
\begin{equation}
\label{eq:a21}
 \frac{d \sigma}{dt_2} (\gamma p \rightarrow J/\psi
+ p) \; \simeq \; 70~{\rm nb} \left ( \frac{W}{100~{\rm GeV}}
\right )^{0.83} \: e^{bt_2},
\end{equation}
with slope $b \simeq 4.5~{\rm GeV}^{-2}$, practically independent
of the centre-of-mass energy $W$.

To a good approximation, $s$-channel helicity is conserved in the
$\gamma p \rightarrow J/\psi + p$ process and, for small
$q_{1t}^2$, the amplitudes ${\cal M}_{\lambda_\gamma,
\lambda_\psi}$ satisfy
\begin{equation}
\label{eq:a22}
 \left | \frac{{\cal M}_{00}}{{\cal M}_{11}} \right |^2 \; \sim \;
 {\cal O} \left ( \frac{q_{1t}^2}{M_\psi^2} \right ).
\end{equation}
Thus, for the small values of $q_{1t}^2$ of interest here, it is
safe to neglect production via longitudinally polarized photons.

Now the full calculation of the differential cross section for the
exclusive process $pp \rightarrow p + J/\psi + p$ via photon
exchange goes well beyond the equivalent photon approach described
in Section~2.  First we must allow for the proper treatment of the
photon polarisation as discussed in Section~3.  Then we must
include the amplitudes of inverted diagrams, analogous to those in
Fig.~1 but with the photon coupled to the lower proton.  Fig.~4
shows the final result for a realistic RHIC configuration.  The
effect of rescattering is clearly pronounced in the region $-t
\sim 0.1~{\rm GeV}^{-2}$, and displays a rich $\phi$ dependence.
Recall, $\phi$ is the azimuthal angle between the transverse
momenta, $\vec{p}_{1t}$ and $\vec{p}_{2t}$, of the outgoing
protons.  The cross section is sizeable.  For example, in a
typical bin, $\Delta p_{1t}^2 \sim 0.02~{\rm GeV}^2, \Delta p_{2t}^2
\sim 0.2~{\rm GeV}^2$ and $\Delta \ln \xi_1 \sim 1$, we expect
more than 0.1~pb for $-t \sim 0.1~{\rm GeV}^2$.  However the
$J/\psi \rightarrow \mu^+ \mu^-$ branching ratio has not been
included.

\begin{figure}[!h]
\begin{center}
\epsfig{figure=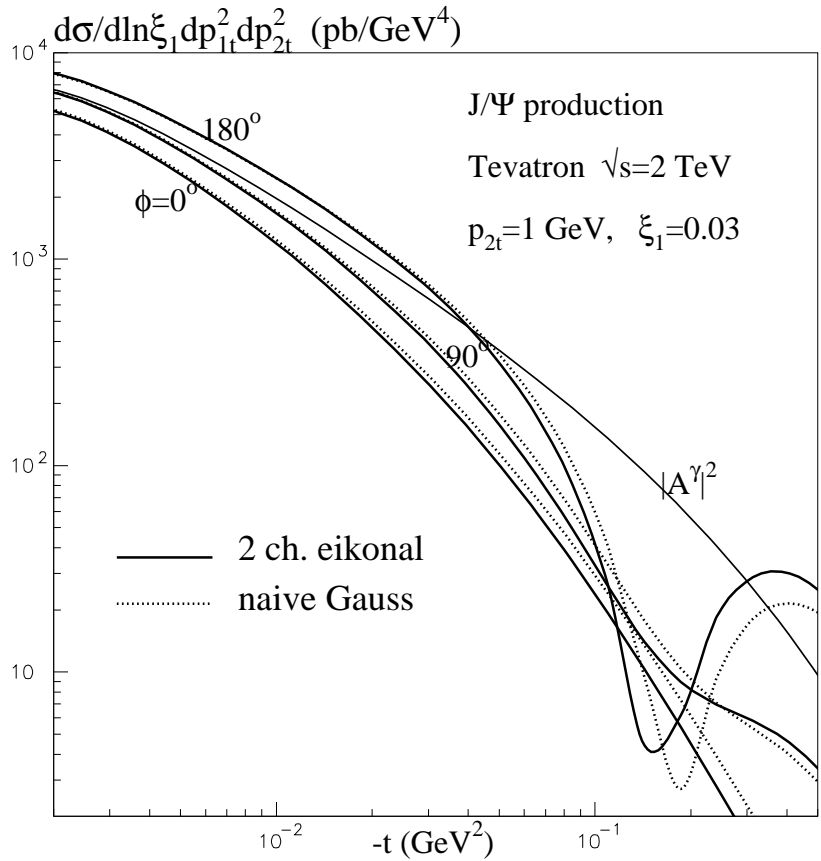,height=5in}
 \caption{As in Fig.~3, but for $p\bar{p} \rightarrow p + J/\psi + \bar{p}$
 at $\sqrt{s} = 2$~TeV.  The dotted curves correspond to the simple $pp$
 rescattering amplitude described in Section~2.}
 \label{Fig5}
\end{center}
\end{figure}

In Fig.~5 we present the analogous results for $p\bar{p}$
collisions\footnote{Note that for $p\bar{p}$ collisions the
`inverted' diagrams enter with the opposite sign.} at the Tevatron
energy $\sqrt{s} = 2$~TeV. To demonstrate the sensitivity of the
differential cross section to the profile of the proton opacity,
we compare the results obtained using the two-channel eikonal
model \cite{KMRsoft}, described in Section~4, with the values
calculated using the simple $pp$ Gaussian amplitude of
(\ref{eq:a7}) with the same slope $B = 17~{\rm GeV}^{-2}$ and
$\sigma_{pp}^{\rm tot} = 76$~mb that appear in the two-channel eikonal model.
Unfortunately, for the reasons we now discuss, it will be
difficult to observe $p\bar{p} \rightarrow p + J/\psi + \bar{p}$
at the Tevatron.

\section{Observability of $pp \rightarrow p + J/\psi + p$}

It is clearly important to measure the kinematic variables
$\vec{p}_{it}$ and $x_i = 1 - \xi_i$ of the outgoing protons.  To
study photon-exchange reactions we require a proton at very small
transverse momentum, say $\vec{p}_{1t}$.  Unfortunately it is not
possible to measure very small $\vec{p}_{it}$ and $\xi_i$
simultaneously since, in this case, an outgoing proton scatters into
the beam pipe.  Thus we choose $p_{2t} \sim 1$~GeV/c and $\xi_1 >
0.01$.  Indeed, in practice it might be necessary to have $\xi_1 >
0.05$.  It is also necessary to identify the $J/\psi$ vector
meson, so as to be sure that we have negative $C$ parity
production (and do not replace photon by Pomeron exchange).

\begin{figure}[!h]
\begin{center}
\epsfig{figure=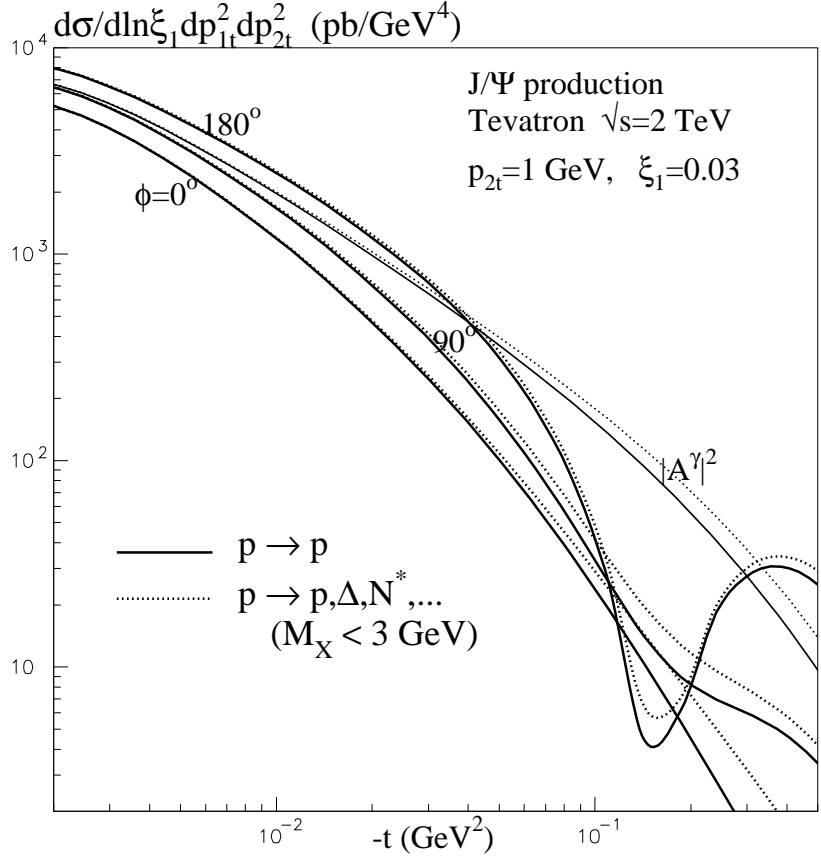,height=5in}
 \caption{As in Fig.~5, but here the dotted curves allow for excitations
 up to mass 3~GeV of the unobservable proton.}
 \label{Fig6}
\end{center}
\end{figure}

In principle, with very good resolution, it would be possible to
observe $J/\psi$ as a peak in the missing mass spectrum.  In
practice, this will be extremely difficult.  Thus we need to
observe the decay $J/\psi \rightarrow \mu^+ \mu^-$.  (Note that
the branching fraction for this decay is not included in the
results that we present.)~~For Tevatron energies, however, with
$\xi_1 = 0.05$, the muons are emitted at very small centre-of-mass
angles, less than $3^\circ$.  Therefore in Section~3 we presented
predictions for RHIC energies, where the situation is better.

Another possibility is to measure only one recoil proton with
$p_{2t} \sim 1$~GeV/c and to observe the $J/\psi$.  Then
$\vec{p}_{1t}$ may be deduced as $-(\vec{p}_{2t} +
\vec{p}_{\psi t})$.  However there may be some inelastic contribution
from processes in which the undetected proton is knocked into an
excited state.  This contribution can be calculated when the upper
limit for the mass of the excitation is specified.  Qualitatively,
the contribution of the excited states acts as if it were
described by a spin-flip amplitude, which does not interfere with
the main non-flip contribution.

To demonstrate the possible effect, we show in Fig.~6 the
prediction which includes the excitation of the unobservable
proton into $\Delta$ and $N^*$ resonances up to mass $M = 3$~GeV.
We see that the additional contribution does not spoil the rich
$t$ and $\phi$ dependence of the differential cross section.

There is a potential background for $pp \rightarrow p + J/\psi +
p$ process coming from double-Pomeron $\chi$ production, $pp
\rightarrow p + \chi + p$ with $\chi \rightarrow J/\psi + \gamma$.
The cross section for $\chi_c(0^{++})$ production at the Tevatron has been
estimated to be $d\sigma/dy \sim 120~$nb \cite{KMRmm}.  That is
$d\sigma/dy dp_{1t}^2 dp_{2t}^2 \sim 20-40~{\rm nb/GeV}^4$ at $-t
\sim 0.1~{\rm GeV}^2$.  The $\chi (0^{++}) \rightarrow J/\psi +
\gamma$ branching fraction is 0.007, so the background is a few
times larger than the signal.  Thus, with a $J/\psi$ mass
resolution no better than 0.4~GeV, it is necessary to observe the
$\gamma$ from $\chi$ decay.  The $\chi (2^{++})$ and $\chi
(1^{++})$ states have much larger $J/\psi + \gamma$ branching
fractions, but much smaller production cross
sections and give less background than $\chi (0^{++})$.

\section{$J/\psi$ production via odderon exchange}

In principle, the process $pp \rightarrow p + J/\psi + p$ may also
be mediated by odderon exchange \cite{LN}.  That is the photon in
Fig.~1 may be replaced by a three-gluon $t$-channel state, where
each pair of gluons form a symmetric colour octet, $8_s$.  Such a
three-gluon state has negative $C$-parity and describes odderon
exchange.  Little is known about the odderon amplitude.  So far,
the odderon has not been seen experimentally.  There are indications
that the odderon-nucleon coupling is small \cite{CKMS}.  In
particular, the coupling is zero in the specific model in which
the nucleon is made up of a quark and a (point-like) diquark.

At leading $\alpha_S \ln s$ order the intercept of the odderon
trajectory is predicted to be very close to 1 \cite{BLV}.  From
this point of view a single three-gluon exchange amplitude appears
as a natural model for the odderon.  For the quark-(charm)quark
interaction, shown in Fig.~7, the amplitude is
\begin{equation}
\label{eq:a23}
 T_{qc} \; = \; \frac{10 \alpha_S^3}{81 \pi} \: \int \: \frac{d^2 k_{1t}}{k_{1t}^2}
 \: \frac{d^2 k_{2t}}{k_{2t}^2} \: \frac{d^2 k_{3t}}{k_{3t}^2} \:
 \delta ^{(2)} \left ( \vec{q}_t - \vec{k}_{1t} - \vec{k}_{2t} -
 \vec{k}_{3t} \right ).
\end{equation}
For the case of interest, $\xi_1 \ll 1$, we have $k_i^2 \simeq -
k_{it}^2$, $q_t$ is the total momentum transferred by the
odderon.  The numerical factor $10/81\pi$ arises from (i) the
symmetry of the gluons (1/3!), (ii) the sum over the gluon colour
indices $(\sum |d_{abc}|^2 = 40/3)$, (iii) averaging over the
colours of the incoming quarks, and (iv) $1/2 \pi$ from the
Feynman loop integration.

\begin{figure}[!h]
\begin{center}
\epsfig{figure=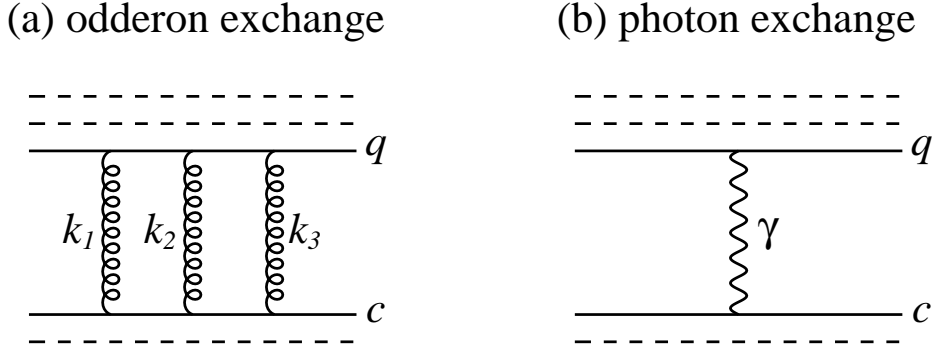,height=2in}
 \caption{The comparison of the (a) odderon- and (b) photon-exchange
 contributions to the quark-(charm)quark amplitude, which are relevant for the
 process $pp \rightarrow p + J/\psi + p$.  The spectator quarks are shown
 by the dashed lines, and all permutations of the gluons coupling to
 different quarks are implied.}
 \label{Fig7}
\end{center}
\end{figure}

Fig~7 is not the only, or necessarily the largest\footnote{There
is also the probability that one of the three $t$-channel gluons
could couple directly to the lower proton rather than to the {\it c} quark
and, together with another {\it t}-channel gluon, forms a Pomeron in the
lower part of the diagram for the $pp \rightarrow p
+ J/\psi + p$ process.}, odderon exchange contribution to $pp
\rightarrow p + J/\psi + p$. However it should give a reasonable
preliminary order-of-magnitude estimate of the possible magnitude
of odderon-exchange.  For such an estimate we take amplitude
(\ref{eq:a23}) with an appropriate infrared cut-off given by the
size of the $J/\psi$ meson; $k_0 = 1/m_c$, and assume quark
additivity.  That is, we assume the coupling of the odderon to the
(upper) proton is three times the odderon-quark coupling.  Then
the odderon-exchange contribution to the proton-charm quark
amplitude is
\begin{equation}
\label{eq:a24}
 T_{pc}^{\rm odderon} \; \sim \; 3 \: \frac{10 \alpha_S^3}{81 \pi}
 \left ( \frac{15 \pi^2}{m_c^2} \right ) \; \sim \; 1~{\rm
 GeV}^{-2},
\end{equation}
using $\alpha_S \sim 0.5$, as compared to the photon-exchange
contribution
\begin{equation}
\label{eq:a25}
 T_{pc}^\gamma \; = \; \frac{4 \pi \alpha}{q_t^2} \: 2e_c \; = \;
 \frac{0.12}{q_t^2}
\end{equation}
with the charge of the charm quark $e_c = 2/3$.  The expression in
brackets in (\ref{eq:a24}) is the estimate of the integral in
(\ref{eq:a23}) with cut-offs $k_{it} > m_c$.\footnote{Technically,
for each $k_{it}$ below ${\it m_c}$, the integrand in (\ref{eq:a23}) is multiplied
by an additional factor $k_{it}/m_c$.}  Comparing
(\ref{eq:a24}) and (\ref{eq:a25}), it is evident that photon
exchange dominates for very small $q_t^2$, but already by $q_t^2
\sim 0.1~{\rm GeV}^2$ odderon-exchange may become of comparable
importance.

It may be possible to use $\phi$ or $\Upsilon$ vector meson
production to distinguish between odderon and $\gamma$ exchange.
If we compare $\Upsilon$ to $J/\psi$ production, then see that the
odderon amplitude is suppressed by $1/m_b^2$ as compared to $1/m_c^2$.
However the $pp \rightarrow p + \Upsilon + p$ rate via
photon-exchange, times the $\Upsilon \rightarrow \mu^+ \mu^-$
branching fraction, is more than three orders of magnitude smaller
than for $J/\psi$ production, and so the signal will be difficult
to observe.

On the other hand, the cross section for $\phi$ production, via
photon-exchange, exceeds $J/\psi$ by a factor of 20, and so it
should be possible to observe the $\phi \rightarrow K\bar{K}$ and
$\mu^+ \mu^-$ decay modes.  Here the relative magnitude of the
photon-exchange amplitude is
suppressed by the strange quark charge $(e_s = - 1/3)$, while the
odderon is enhanced by the large size of the $\phi$ meson.
Therefore exclusive $\phi$ production appears to be a promising
way to search for the odderon.

Finally, intrinsic charm \cite{BROD} is another potential
mechanism for the process $pp \rightarrow p + J/\psi + p$. However
it is unlikely that there is enough intrinsic charm component in
the proton at small $\xi_1$.  The mechanism is to replace photon
exchange by a $c\bar{c}$ pair in the $t$-channel.  Such an
amplitude dies out linearly with $\xi_1$ (modulo logarithmic
corrections) as compared to photon-exchange contribution.  The
lowest order perturbative QCD estimate gives a
contribution an order-of-magnitude smaller than odderon exchange.

\section{$W$ boson hadroproduction via $\gamma$ exchange}

Another possible way to investigate rescattering effects is
central $W$ production in the process $pp \rightarrow p + W + X$,
where the $W$ boson is separated from the dissociating proton by a
large rapidity gap, $\Delta \eta$.  The leading contribution comes
from the diagram shown in Fig.~8a.  Other configurations where the
exchanged photon interacts with a quark emitting the $W$ boson,
rather than directly with the $W$, are suppressed by a
factor $\exp (-\Delta \eta)$.  (A
possible odderon-exchange contribution would be suppressed by the
same factor, as gluons cannot couple to the $W$ boson.)~~Our
predictions below are based on the leading configuration.

\begin{figure}[!h]
\begin{center}
\epsfig{figure=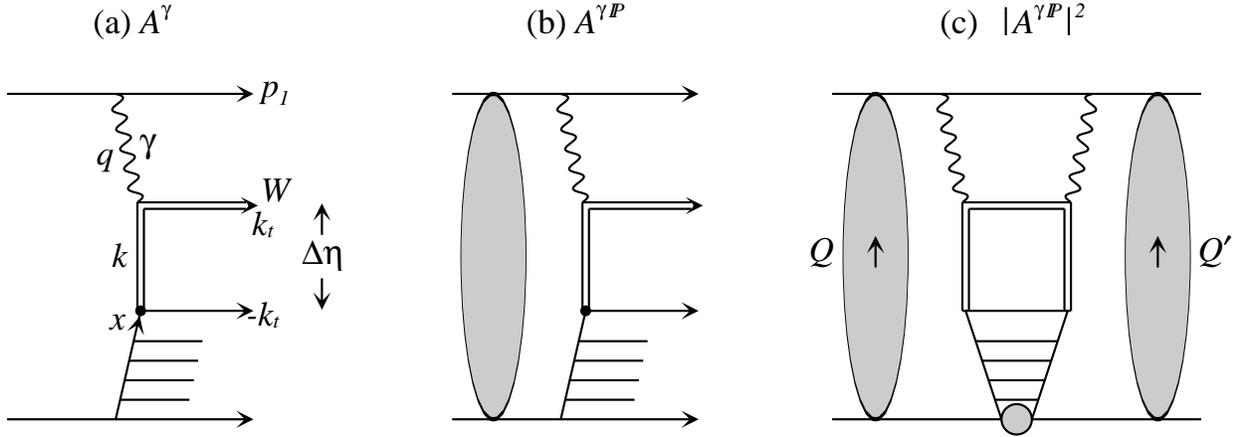,height=2.5in}
 \caption{Photon-exchange amplitude for the process $pp \rightarrow
 p + W + X$, (a) without, and (b) with, rescattering effects.  Diagram
 (c) shows the rescattering amplitude times its complex conjugate.}
 \label{Fig8}
\end{center}
\end{figure}

The cross section corresponding to the diagram of Fig.~8(a) may be
written as
\begin{equation}
\label{eq:a26}
 d \sigma \; = \; dN^T \sigma (\gamma p \rightarrow W + X),
\end{equation}
where the effective number of photons $N^T$ is given by
(\ref{eq:a4}), and
\begin{equation}
\label{eq:a27}
 \sigma (\gamma p \rightarrow W + X) \; = \; \int_{x_{\rm min}}^1
 \: \frac{dx}{x} \: U (x) \: \frac{g^2 \alpha}{4} \: \frac{(1 + k_t^2/2M_W^2)}{(M_W^2 +
 k^2)^2} \: dk_t^2,
\end{equation}
where $k_t^2$ is the square of the momentum transferred through
the $t$-channel $W$ boson and $g^2 = 8 M_W^2 G_F/\sqrt{2}$.  The
density of quarks which may emit the $W^+$ boson is
\begin{equation}
\label{eq:a28}
 U (x) \; = \; xu (x, k_t^2) \: + \: x\bar{d} (x, k_t^2) \:
 + \: x \bar{s} (x, k_t^2) \: + \: xc (x, k_t^2).
\end{equation}
For $W^-$ production, $U
(x)$ is replaced by
\begin{equation}
\label{eq:a29}
 D (x) \; = \; x\bar{u} (x, k_t^2) \: + \: xd (x,
k_t^2) \: + \: xs (x, k_t^2) \: + \: x \bar{c} (x, k_t^2).
\end{equation}
The minimal momentum fraction $x_{\rm min}$ carried by the parent
quark is limited by the requirement that the recoil quark jet with
transverse momentum $k_t$ be outside the rapidity gap $\Delta
\eta$, that is\footnote{In the case of recoil charm-quark jet we substitute $k_t$
in the second term in (\ref{eq:a30}) by $\sqrt{k_t^2 + m_c^2}$.}
\begin{equation}
\label{eq:a30}
 x_{\rm min} \; = \; \sqrt{\frac{M_W^2 + k_t^2}{s}} \: \exp (-y) \:
 + \: \frac{k_t}{\sqrt{s}} \: \exp (-y + \Delta \eta),
\end{equation}
where $y = \frac{1}{2} \ln \left ((E + k_z)/(E - k_z) \right )$ is
the $W$ boson rapidity.  The two terms in the expression in
brackets in the numerator in (\ref{eq:a27}) correspond to the
production of transversely and longitudinally polarised $W$
bosons.

The amplitude with rescattering $A^{\gamma \funp}$, Fig.~8(b), has
a form similar to (\ref{eq:a20}).  Again we allow for photon
polarisation effects.  However, now there is almost no correlation
between $Q_t$ and $k_t$, as the relatively small loop momentum
$Q_t$ is separated from the $W$ transverse momentum $k_t$ by a long
evolution chain.  The only novel point is that in the rescattering
term squared, $|A^{\gamma \funp}|^2$, the loop momenta $Q_t$ and
$Q_t^\prime$ (corresponding to the complex conjugate amplitude)
are correlated via Fig.~8(c) by the form factor $\exp \left (-b_N
(\vec{Q}_t + \vec{Q}_t^\prime)^2 \right )$.  The slope $b_N$,
associated with the proton-Pomeron vertex, characterises the
spatial distribution of the quarks in the proton before the DGLAP
evolution.  Note that for $pp \rightarrow p + W + X$, as one
proton dissociates, there are no complications from `inverted'
diagrams.

\begin{figure}[!h]
\begin{center}
\epsfig{figure=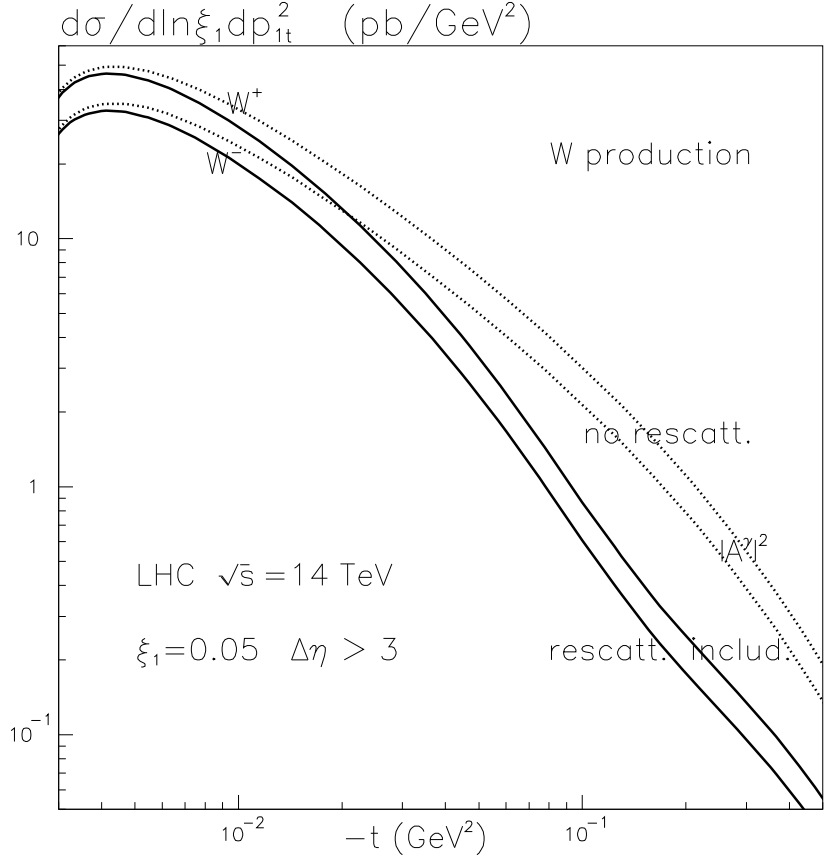,height=5in}
 \caption{The differential cross section for $pp \rightarrow p + W^\pm + X$
 at the LHC.  The dotted and continuous curves correspond, respectively, to the
 predictions without and with the rescattering effects of Figs.~8(b,c).  In each
 case $W^+$ production corresponds to the upper one of the pair of curves.
 The rapidity gap between the quark recoil jet and the $W$ boson is taken to
 satisfy $\Delta \eta > 3$.}
 \label{Fig9}
\end{center}
\end{figure}

Sample results for the $pp \rightarrow p + W^\pm + X$ differential
cross sections at the LHC energy are shown in Fig.~9.  The dotted
and continuous curves correspond to the results without, and with,
rescattering effects respectively.  As the azimuthal angle $\phi$
has been integrated over the diffractive dip, which was clearly
visible at $\phi = 180^\circ$ for $pp \rightarrow p + J/\psi + p$,
is now very shallow.  We see that rescattering suppresses the
cross section by a factor of about 4 for $-t \sim 0.1~{\rm
GeV}^2$.  The rescattering is calculated using the prescription of
Ref.~\cite{KMRsoft}.  If the more complicated model of
Ref.~\cite{KKMR} is used, then we find the predictions change by
less than 10\%, the effect is maximal in the region of the dip
($|t| \sim 0.1$ GeV$^2$), where the rate rises by about 10\%.

Of course, the process is hard to observe in the $W \rightarrow
q\bar{q}$ decay mode, due to the huge QCD background.  However, the
cross section is large enough to be observed in the leptonic decay
modes.  For example, from Fig.~9 we see that the $\Delta p_{1t}^2
\simeq 0.03~{\rm GeV}^2, \Delta \ln \xi_1 \simeq 1$ bin at $-t
\sim 0.1~{\rm GeV}^2$ has a cross section of about 20~fb.  If the
momenta of the charged lepton, the outgoing proton and the recoil
jet can be measured, then we can reconstruct the mass of the $W$
boson.

\section{$\mu^+ \mu^-$ pair production via $\gamma\gamma$ fusion}

The exclusive process $pp \rightarrow p + (\mu^+ \mu^-) + p$
proceeds via $\gamma\gamma$ fusion.  Again we avoid the odderon or
$q\bar{q}$ exchange mechanisms.  The QED $\gamma\gamma \rightarrow
\ell^+ \ell^-$ cross section can be calculated to good accuracy,
and if we select events with very small $p_{1t}$ of the leading
proton, then the process may even be used to measure the incoming
$pp$-luminosity \cite{st,KMRlum}.  On the other hand, as $p_{1t}$
increases the reaction becomes sensitive to the rapidity gaps
survival factor $\hat{S}^2$.  Hence, by varying the transverse
momentum $p_{1t}$ we may scan the proton-proton opacity $\Omega
(b_t)$.  The experimental problem is that to obtain a sufficient
event rate we need to identify leptons (either muons or electrons)
with transverse energy, $E_T$, as small as 1 or 2~GeV and large
rapidity\footnote{To measure small $p_{1t} \sim 100$~MeV we
require $\xi_1 \gapproxeq 0.03$.  For the LHC it means that the
longitudinal momentum of the lepton is $p_{|\!|} \sim \xi_1
\sqrt{s}/4 \sim 100$~GeV.  Thus the rapidity $\eta = -\ln \tan
\theta/2 \simeq \ln (2p_{|\!|}/E_T) \sim 5$, and the polar angle
of the lepton $\theta \simeq E_T/p_{|\!|} < 0.02$.},
$\eta \gapproxeq 5$.

To illustrate the size of the effect we use a simplified form of
the QED cross section, which corresponds to the limit $E_T \gg
q_{1t}, q_{2t}$ and $m$, where $q_i$ are the photon momenta and
$m$ is the lepton mass.  After averaging over the azimuthal angle
of the transverse momentum $\vec{E}_T$ of the lepton, the $\gamma\gamma
\rightarrow \ell^+ \ell^-$ subprocess amplitude squared is, see ref. \cite{KMRlum},
\begin{equation}
\label{eq:a31}
 \frac{d \hat{\sigma}}{d \hat{t}} \; = \; \frac{\cosh (\Delta \eta)
 \pi \alpha^2}{4 E_T^4 \cosh^4 (\Delta \eta/2)} \: \left (
 \frac{(\vec{q}_{1t} \cdot \vec{q}_{2t}^{~\prime})(\vec{q}_{2t} \cdot \vec{q}_{1t}^{~\prime})
 + (\vec{q}_{1t} \cdot \vec{q}_{1t}^{~\prime})(\vec{q}_{2t} \cdot \vec{q}_{2t}^{~\prime}) -
 (\vec{q}_{1t}^{~\prime} \cdot \vec{q}_{2t}^{~\prime})(\vec{q}_{1t} \cdot \vec{q}_{2t})}{q_{1t}
 q_{1t}^\prime q_{2t} q_{2t}^\prime} \right ).
\end{equation}
As before, we use $q_{it}^\prime$ to denote the photon transverse momenta in the
complex conjugate amplitude.
The relatively complicated form is associated with the
polarisation vectors $\vec{\varepsilon}_i$ of the photons, which
are parallel to the photon transverse momenta.  Instead of a
simple form like $\vec{\varepsilon} \cdot
\vec{\varepsilon}^{~\prime}$, which occurred in $J/\psi$
production, we now face correlations like $(\vec{\varepsilon}
\cdot \vec{E}_T)(\vec{\varepsilon}^{~\prime} \cdot \vec{E}_T)$.
After including all such terms, and averaging over the $\vec{E}_T$
direction we obtain (\ref{eq:a31}).  Recall that in the
rescattering contribution the primed photon momenta $q_{1t}^\prime =
Q_t^\prime - p_{1t}^\prime$ of the complex conjugate amplitude, may
differ from the momenta $q_{1t} = Q_t - p_{1t}$ occurring in the
amplitude (and analogously for $q_{2t}^\prime$), see Fig.~3(a,b) of
Ref.~\cite{KMRlum}. In the absence of rescattering we put $Q = 0$ and/or
$Q^\prime = 0$.  Finally,
the cross section (\ref{eq:a31}) has to be convoluted with the
fluxes (\ref{eq:a4}) for photons of momentum $q_1$ and $q_2$.

Sample results are presented in Fig.~10 for the LHC energy.  The
cross section has been integrated over the mass of the muon pair,
with the cut $E_T > 4$~GeV on each muon.  Again we see a clear
diffractive dip at $-t = 0.2~{\rm GeV}^2$ for the back-to-back
proton configuration, $\phi = 180^\circ$.  Unfortunately the cross
section is small and will make observation difficult.

In analogy to $J/\psi$ production, it may be possible to measure
only recoil proton and to observe the $\mu^+\mu^-$ pair, with
$\vec{p}_{1t}$ reconstructed via $\vec{p}_{2t} +
\vec{p}_{\mu^+\mu^-t}$. Here we can benefit from the high muon
momentum resolution.

\begin{figure}[!h]
\begin{center}
\epsfig{figure=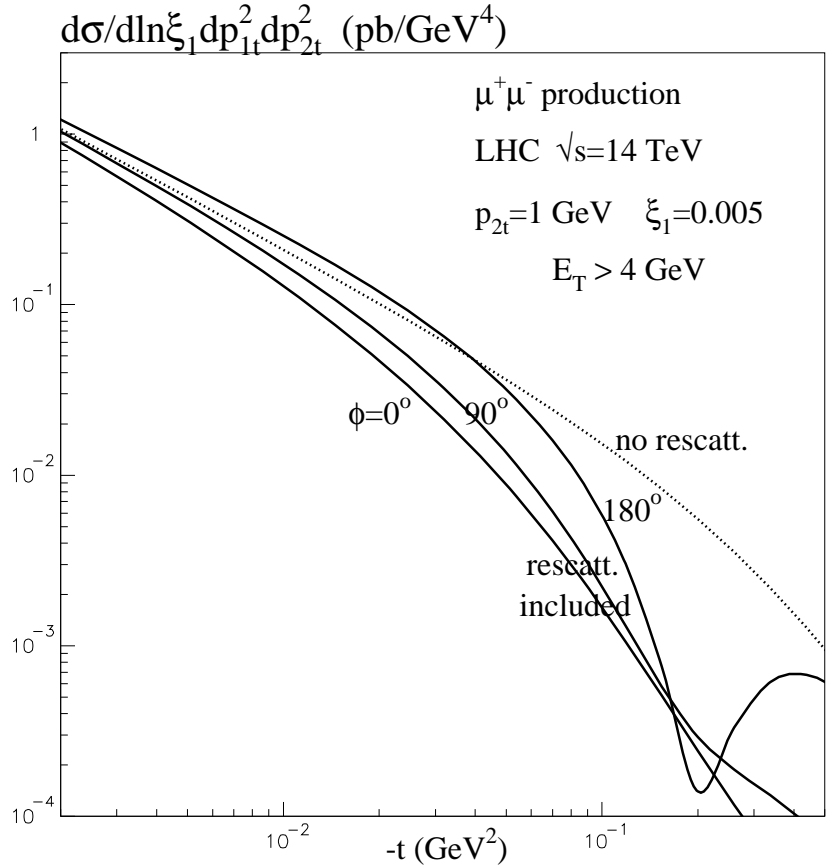,height=5in}
 \caption{The cross section for $pp \rightarrow p + (\mu^+ \mu^-) + p$ at
 the LHC energy, with (continuous curves) and without (dotted curve)
 rescattering effects included.
 The rescattering effects are shown for three values of the azimuthal angle $\phi$
 between the transverse momenta, $\vec{p}_{1t}$ and $\vec{p}_{2t}$, of the
 outgoing protons.}
 \label{Fig10}
\end{center}
\end{figure}

\section{Conclusions}

We have demonstrated that $\gamma$-exchange processes at hadron
colliders (such as $pp \rightarrow p + M + p)$ provide a good
testing ground for checking our ability to calculate the effects
of rescattering, and in this way to scan the survival probability
of rapidity gaps.  We have taken the centrally produced system to
be either the $J/\psi$ vector meson or the $W$ boson or a $\mu^+
\mu^-$ pair.  We have emphasized the important role played by
photon polarisation; it may even reverse the sign of the
absorptive corrections for small $t$ and $\phi = 180^\circ$ (where
$\phi$ is the azimuthal angle between the transverse momenta of
the two outgoing protons).

The interference between the pure $\gamma$ exchange amplitude
$A^\gamma$ and the amplitude $A^{\gamma \funp}$ with rescattering
effects generates a rich diffractive structure of the differential
cross section for processes of the type $pp \rightarrow p + M +
p$.  The tagging of both the leading protons, together with the
centrally produced system, allows the rich $t$ and $\phi$
dependence of the cross section to be measured.  In this way the
optical density of the proton-proton rescattering interaction can
be probed as a function of the impact parameter.

For LHC energies, $W$ boson production with a rapidity gap, looks
the most promising probe of the gap survival probability.  On the
other hand, the $J/\psi$ production process looks more realistic
at RHIC energies (provided the background from $\chi \rightarrow J/\psi+\gamma$
can be supressed).  Indeed a comparison of $J/\psi$ and $\phi$
meson production also offers an attractive opportunity to search
for odderon-exchange.

\section*{Acknowledgements}

We thank Risto Orava and Krzysztof Piotrzkowski for interesting
discussions. One of us (VAK) thanks the Leverhulme Trust for a
Fellowship. This work was partially supported by the UK Particle
Physics and Astronomy Research Council, by the Russian Fund for
Fundamental Research (grants 01-02-17095 and 00-15-96610) and by
the EU Framework TMR programme, contract FMRX-CT98-0194 (DG
12-MIHT).

\end{document}